\documentclass[prl,aps,showpacs,superscriptaddress,twocolumn]{revtex4}
\usepackage{color} 
\usepackage{bm}
\usepackage{dcolumn}
\usepackage{times}
\usepackage{amssymb} 
\usepackage{amsmath} 
\usepackage{ragged2e}
\usepackage{graphicx}
\usepackage{epstopdf}
\usepackage[english]{babel}
\usepackage{mathrsfs}
\usepackage{textcomp}
\usepackage{amsthm}
\usepackage{amsfonts}
\usepackage{soul}
\usepackage{braket} 
\usepackage{hyperref}
\usepackage{cleveref}
\newcommand{\be}{\begin{equation}}
\newcommand{\ee}{\end{equation}}
\newcommand{\bea}{\begin{eqnarray}}
\newcommand{\eea}{\end{eqnarray}}
\newcommand{\ba}{\begin{array}}
\newcommand{\ea}{\end{array}}

\begin{document}

\title{Quenched dynamics and spin-charge separation in an interacting topological lattice}
\author{L. Barbiero}
\affiliation{Center for Nonlinear Phenomena and Complex Systems,
Universit\'e Libre de Bruxelles, CP 231, Campus Plaine, B-1050 Brussels, Belgium}
\author{L. Santos} \affiliation{Institut f\"ur Theoretische Physik,
  Leibniz Universit\"at Hannover, Appelstr. 2, DE-30167 Hannover,
  Germany}
\author{N. Goldman}
\affiliation{Center for Nonlinear Phenomena and Complex Systems,
Universit\'e Libre de Bruxelles, CP 231, Campus Plaine, B-1050 Brussels, Belgium}

\begin{abstract}

We analyze the static and dynamical properties of a one-dimensional topological  lattice, the fermionic Su-Schrieffer-Heeger model, in the presence of on-site interactions. Based on a study of charge and spin correlation functions, we elucidate the nature of the topological edge modes, which depending on the sign of the interactions, either display particles of opposite spin on opposite edges, or a pair and a holon. This study of correlation functions also highlights the strong entanglement that exists between the opposite edges of the system. This last feature has remarkable consequences upon subjecting the system to a quench, where an instantaneous edge-to-edge signal appears in the correlation functions characterizing the edge modes. Besides, other correlation functions are shown to propagate in the bulk according to the light-cone imposed by the Lieb-Robinson bound. Our study reveals how one-dimensional lattices exhibiting entangled topological edge modes allow for a non-trivial correlation spreading, while providing an accessible platform to detect spin-charge separation using state-of-the-art experimental techniques.
\end{abstract}


\maketitle

Topological phases of matter exhibit unusual quantum properties~\cite{hasan,qi}, which are currently investigated in a wide range of physical platforms~\cite{Asorey,Goldman_topology,Ozawa_review}. While the traditional quantum Hall effects~\cite{klitzking,laughlin,thouless,moore} and topological insulators~\cite{kane,bernevig} are found in two-dimensional (2D) or three-dimensional (3D) materials, special attention has been recently devoted to the study of one-dimensional (1D) systems with topological features~\cite{kitaev,Jiang_2011,Alicea_2011,kraus,verbin,ryu,deplace,diliberto,gorlach,Christina_Marcello,Alejandro,icfo_dynamical,sbierski}. A prominent example is provided by the Su-Schrieffer-Heeger (SSH) model~\cite{su}, which belongs to the class BDI of 1D chiral Hamiltonians~\cite{hasan}, and which offers a minimal setting for the study of non-trivial topology, robust boundary states and charge fractionalization~\cite{heeger,jackiw}. The simplicity and richness of this toy-model, which was originally introduced to describe doped polyacetylene, strongly motivated its recent experimental implementation, both in ultracold bosonic gases~\cite{atala,meier} and photonics~\cite{Malkova2009,stjean,Ozawa_review}. Until now, such experiments operated in the non-interacting regime, where topological properties are thus fully understood at the single-particle level.

In the presence of inter-particle interactions, 1D quantum systems generically show striking manifestations of genuine quantum-mechanical effects ~\cite{giamarchi}, hence ruling out any semiclassical description. In this context, the Tomonaga-Luttinger theory~\cite{tomonaga,luttinger} provides accurate predictions for low-energy excitations. The most surprising result emanating from this theory is the well-known phenomenon of spin-charge separation, which reflects the fact that spin and charge excitations can behave independently and move at different speeds. While an experimental demonstration of this effect has been reported~\cite{kim,jompol}, one still lacks a stable platform where spin-charge separation can be studied in a clean and systematic manner.

Besides, the out-of-equilibrium dynamics of many-body quantum systems generally exhibit the Lieb-Robinson locality phenomenon, which constrains information to propagate through a system with a finite bounded velocity~\cite{lieb,nachtergaele}, hence manifesting in a light-cone signal spreading. Its deep connection to fundamental principles of quantum mechanics~\cite{hastings1,hastings2,eisert}, such as thermalization~\cite{rigol,gong}, information propagation in quantum channels~\cite{bose}, entanglement scaling~\cite{eisert,hastings3} and correlation decay~\cite{hastings2,bravyi}, strongly motivated the experimental demonstration of this concept in ultracold bosonic gases \cite{cheneau,langen} and trapped ions \cite{richerme,jurcevic}. Moreover, it has been recently suggested that out-of-equilibrium dynamics can also reveal unique topological signatures ~\cite{strinati,sacramento,Caio2015,Alessio2015,Caio2016,Hu2016,Wilson2016,Dehghani2016,Unal2016,Wang2017}. The validity of this approach has been experimentally confirmed in systems of ultracold atoms trapped in shaken optical lattices, where dynamical topological phase transitions~\cite{Flaschner2017} and non-trivial winding numbers~\cite{Tarnowski2017} have been measured.

In this work, we reveal an intriguing interplay between topology, spin-charge separation, and correlations spreading, which is shown to appear in interacting 1D fermionic lattices. Motivated by its simplicity and experimental accessibility, we focus our study on the interacting fermionic SSH model, which we analyze both from a static and dynamical perspective. We start by establishing the nature of the boundary modes by means of correlation functions;  depending on the sign of the interaction, these modes can be either constituted of one up component on one edge and one down component on the other, or of a holon on an edge and a pair (up-down) on the other. Importantly, both configurations are shown to exhibit entanglement between opposite edges. The latter entanglement property has strong consequences upon quenching the system, locally or globally, as it allows for an instantaneous edge-to-edge correlation signal related to the spin or the particle density. Noticeably, an additional bulk signal, which verifies the traditional Lieb-Robinson bound, is also present in all the considered correlation functions. The results presented below demonstrate how topological systems exhibiting entangled edge states allow for non-trivial correlation spreading in their (quenched) dynamics, and also designates such 1D fermionic systems as accessible platforms to experimentally detect spin-charge separation.

\paragraph{Model.}
The fermionic interacting Su-Schrieffer-Heeger (SSH) model is described by the following Hamiltonian,
\begin{equation}
H=-\sum_{i,\sigma} \left [(J+\delta J(-1)^i)c_{i,\sigma}^\dagger c_{i+1,\sigma}+\text{h.c.} \right ]+U\sum_in_{i\uparrow}n_{i\downarrow},
\label{ssh}
\end{equation}
where $J$ represents the nearest-neighbor tunneling amplitude on the lattice, $U$ is the onsite interaction between two fermions with opposite spin and $c_{i,\sigma}^\dagger(c_{i,\sigma})$ describes the creation (annihilation) of a fermion with spin $\sigma$ in the $i$-site of a chain of length $L$; in the following, we set $J\!=\!\hbar\!=\!1$, which sets our energy and time units. The important feature in Eq.~\eqref{ssh} is the dimerization of the tunneling amplitudes, through the parameter $\delta J$, which sets the topological properties of the band structure: in the non-interacting case, one finds that $\delta J>0$ leads to a vanishing Zak phase, which corresponds to a trivial regime, while for $\delta J<0$, the Zak phase has the value of $\pi$ and degenerate edge modes are present at zero energy. 

The topological properties of the SSH model have also been explored in the presence of finite interactions, both for bosonic~\cite{grusdt} and fermionic~\cite{wang,ye} versions of the model. However, in the fermionic case, it is worth pointing out that topological features were only identified through entanglement properties~\cite{wang,ye}, which are hardly accessible in experiments.

In the following, we shall focus on the half-filled balanced configuration, i.e.~$N_\uparrow=N_\downarrow=L/2$, where a finite $U$ preserves the fully-gapped Peierls dimerization~\cite{peierls}. In this regime, particles form dimers in alternating bonds~\cite{yu}, which signals a broken inversion symmetry, as captured by finite values of the parity order parameter in both charge and spin sectors~\cite{barbiero}. 
 In the present context, the role of the interaction in Eq.~\eqref{ssh} is essentially to reduce from 4-fold to 2-fold the degeneracy of the edge-mode manifold associated with the $U\!=\!0$ configuration, namely 
\begin{align}
&\vert 1 \rangle=(\text{right} : \uparrow, \text{left}: \downarrow), \qquad \vert 2 \rangle=(\text{right}: \downarrow, \text{left}: \uparrow), \label{edge_man}\\
&\vert 3 \rangle=(\text{right}: \varnothing,\text{left}: \uparrow \downarrow), \quad \vert 4 \rangle=(\text{right}: \uparrow \downarrow,\text{left}:  \varnothing),\notag 
\end{align}
where $\varnothing$ denotes a holon, and where right/left refer to the two opposite edges. As illustrated below, this modification of the edge-manifold degeneracy has fundamental consequences both in the static and dynamical properties of the system. In the next paragraph, we first elucidate the nature of the edge modes, for a wide range of interaction strengths, through a correlation-functions study.
 
 \paragraph{Static Properties.}
At finite $U$, any $\delta J\!<\!0$ preserves the presence of degenerate edge modes~\cite{SM}. This is confirmed through the behavior of both the entanglement spectrum \cite{Christina_Marcello,pollmann} and the density distribution upon adding a single particle to the system; see~\cite{SM} for details. In order to fully capture the nature of the edge modes, we performed density-matrix-renormalization-group (DMRG) calculations~\cite{white}, which provide the decay of the following correlation functions:
\begin{equation}
C_j=\langle S^C_0S^C_j\rangle , \qquad S_j=\langle S^S_0S^S_j\rangle ,\label{charge}
\end{equation}
where $S^C_j=\sum_{\sigma}n_{j,\sigma}-1$ and $S^S_j=n_{\uparrow,j}-n_{\downarrow,j}$ refer to the charge and spin degrees of freedom, respectively. More precisely, $C_j$ measures the amount of correlation existing between a holon located at the first site of the lattice ($j\!=\!0$) and a pair $(\uparrow \downarrow)$ placed at some site $j\!>\! 0$ (and similarly between a pair at $j\!=\!0$ and a holon at $j\!>\! 0$); while the quantity $S_j$ measures the correlations between a spin-up (resp.~a spin-down) fermion at $j\!=\!0$ and a spin-down (resp.~a spin-up) fermion at $j\!=\!L$.
\begin{figure}
\includegraphics[scale=0.25]{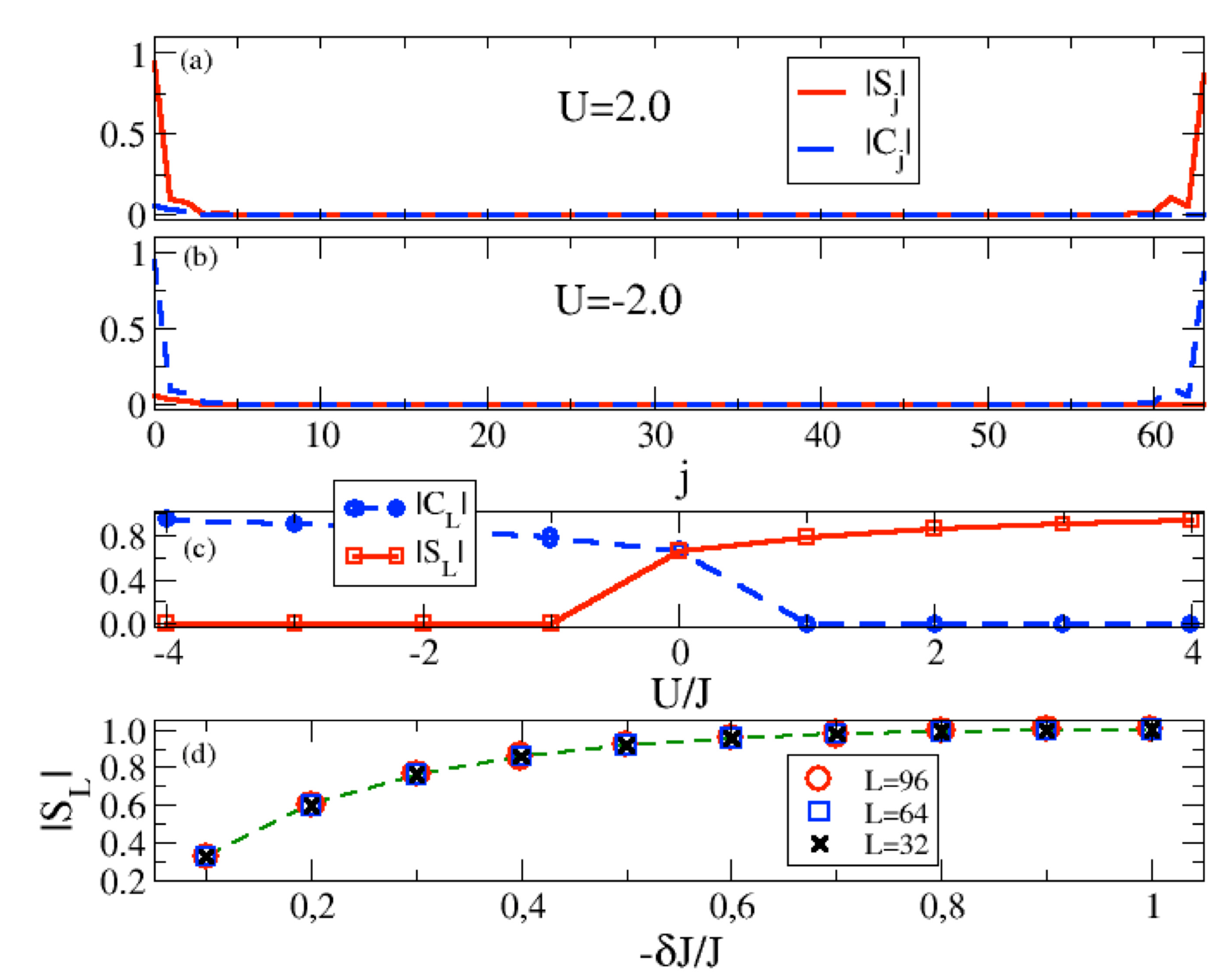}
\caption{(Color online) (a)-(b) Decay of $|S_j|$ and $|C_j|$ for $U=2$ and $\delta J=-0.4$ in a system of length $L=64$. (c) $|S_L|$ and $|C_L|$ as a function of $U$ for $\delta J=-0.4$ in a system of length $L=64$. (d) $|S_L|$ and $|C_L|$ as a function of $\delta J$ for different system sizes $L$ and $U\!=\!2$. All the results are obtained by means of DMRG simulation keeping up to 512 DMRG states and performing 5 finite size sweeps.}
\label{corre}
\end{figure}

As illustrated in Fig.~\ref{corre}(a)-(b), the absolute value of $C_j$ and $S_j$ shows very different behaviors depending on the sign of $U$. For repulsive interactions $U\!>\!0$, both correlations are found to vanish in the bulk, however, $\vert S_j \vert$ shows a large value at the boundary $j\!=\!L$ of the system. Due to the particle-hole symmetry inherent to the model in Eq.~\eqref{ssh}, an identical behavior occurs for $U\!<\!0$, in which case it is the correlation function associated with the charge, $C_j$, which exhibits a finite edge-to-edge signal, see Fig.~\ref{corre} (b). These results allow one to properly capture the nature of the edge modes, but also to characterize the explicit form of the Peierls dimerization, see \cite{SM}. Indeed, for $U\!>\!0$,  the behavior of $|S_j|$ implies that the edge modes are constituted of one up and one down component on opposite edges [i.e.~the states $\vert 1,2 \rangle$ in Eq.~\eqref{edge_man}]; in contrast, for $U\!<\!0$, the finite value of $|C_L|$ at the edges indicates that the edge modes are formed by a holon and a pair of fermions at opposite edges [i.e.~the states $\vert 3,4 \rangle$ in Eq.~\eqref{edge_man}]. This also allows one to conclude that the alternating bonds appearing in the bulk of the system are formed by dimers composed of two fermions (up and down) for $U\!>\!0$, while they are formed by a pair and a holon for $U\!<\!0$; this is also in agreement with energetic considerations. We note that the similar dimerization structure, occurring at $U>0$, is found in an extended Hubbard model~\cite{nakamura}. Moreover, the additional information encoded in Figs.~\ref{corre}(c)-(d) indicates that the localization strength of the edge modes both depends on the interaction strength $U$ (which affects the wave-function overlaps), and on the parameter $\delta J$. 

It is worth emphasizing that for the non-interacting case, $U\!=\!0$, both $|C_L|$ and $|S_L|$ have the exact same value [Fig.~\ref{corre} (c)]. This latter result is in agreement with the aforementioned 4-fold degeneracy of the edge modes [Eq.~\eqref{edge_man}]. One should note that, in practice, this result would correspond to an average over a series of measurements, since monitoring a single edge state could lead to a finite edge contribution to the spin or  charge correlation functions. As explained above, a finite interaction $U\!\ne\!0$ then reduces the system's degeneracy and, depending on its sign, selects the two lowest-energy states [$\vert 1,2 \rangle$ or $\vert 3,4 \rangle$ in Eq.~\eqref{edge_man}], thus giving rise to the distinct edge-to-edge behaviors of the correlations [Fig.~\ref{corre}(a)-(b)]. 

Moreover, the result in Fig.~\ref{corre}(d) suggests that special attention should be paid to the case $\delta J\!=\!-J$, where $|S_L|\!=\!1$. In this peculiar configuration of the dimerization strength, the two edge modes are found to be totally disconnected from the rest of the system, which leads to a vanishing of the correlation length associated with $S_j$, hence giving rise to the special value $|S_L|\!=\!1$. 

As a final remark on static properties, we verified that the long-distance entanglement that exists between the two opposite edges of the system is truly a feature of the boundaries, as it is found to be absent in the bulk; this is in direct analogy with the behavior previously discovered in the context of a dimerized Heisenberg chain~\cite{campos}. Here, we confirmed the absence of bulk-bulk or bulk-edge correlations by observing the trivial character of the correlation functions $\langle S^C_{l}S^C_j\rangle$ and $\langle S^S_{l}S^S_j\rangle$ for $l\!\ne\!0$; we also verified that the finite edge-to-edge correlations are size independent [Fig.~\ref{corre}(d)]. This striking edge-to-edge entanglement could be revealed dynamically in experiments, upon subjecting the system to a quench, as we now discuss in the next paragraph.
 
\paragraph{Dynamical Properties.}
In systems with short-ranged couplings, both local and global perturbations reflect in a light-cone spreading of the correlation functions set by the Lieb-Robinson bound~\cite{lieb,nachtergaele,hastings1,hastings2,eisert,cheneau,langen}. 
While quenched dynamics of topological systems exhibiting degenerate edge modes has been previously studied, in particular to highlight the fragility of topological properties~\cite{strinati,sacramento}, the role of entangled edge modes in the spreading of correlations remains an unexplored topic. In order to capture and describe such a phenomenon, we exploit a time-dependent-density-matrix-renormalization-group (t-DMRG) method~\cite{white1} to study the time-evolution of the equal-time correlation functions in Eq.~\eqref{charge}, $C_j (t)$ and $S_j (t)$, upon subjecting the system to a quench. As a first protocol, we determine the Hamiltonian's ground state, for given $U$ and $\delta J$, and we then let the system evolve after applying a local chemical potential $hS_0^S\!=\!h(n_{\uparrow,0}-n_{\downarrow,0})$ at the left boundary~\cite{note}.   
\begin{figure}
\includegraphics[scale=0.24]{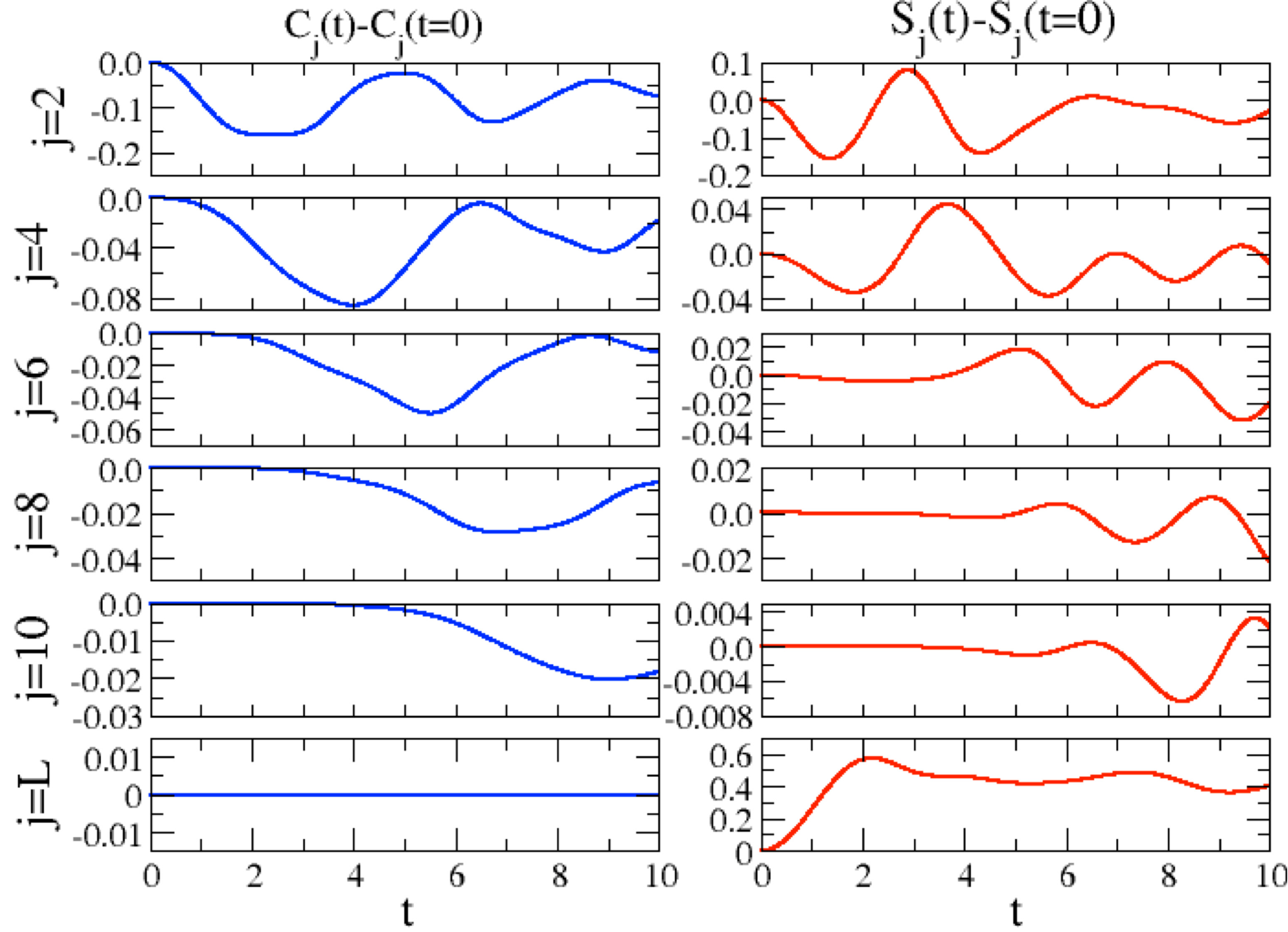}
\caption{(Color online) Spreading of the correlation $C_j(t)-C_j(t=0)$ and $S_j(t)-S_j(t=0)$ for $U=2.0$ and $\delta J=-0.4$ applying a local chemical potential $hS_0^S$ at the first lattice site, i. e. $j\!=\!0$, with $h\!=\!1$. All the results refer to a chain of length $L=40$ obtained by means of t-DMRG simulations keeping up to 512 DMRG states both for the static and for the dynamic and using time step $\delta t=0.01$. }
\label{correDl}
\end{figure}

Fig. \ref{correDl} shows the time-renormalized behavior of $C_j(t)$ and $S_j(t)$ for different values of $j$, in the case where $U\!>\!0$. As clearly visible on the left column, a well defined light-cone-type propagation occurs in the spreading of $C_j$; this behavior is visible in the evolution of $\text{min} [C_j(t)]$ as one considers increasing values of $j$. Due to numerical limitations, this light-cone behavior is shown up to distances of $j\!=\!10$ lattice sites (the simulation time being too short to detect the light-cone signal reaching $j\!=\!L$). A drastically different behavior is found in the spin correlations $S_j(t)$, which captures the effect of the edge modes when $U\!>\!0$ [see previous paragraph]. Indeed, in addition to a clear light-cone propagation, a strong edge-to-edge signal is detected in $S_L(t)$ at short time $t$ [see last panel of Fig.~\ref{correDl}]. Based on our knowledge of the system's static properties, we attribute this quasi-instantaneous correlation spreading between the first and last lattice sites to the entanglement characterizing the spin-up and spin-down states that are localized at the system's boundaries. Besides, the fact that spin and charge excitations propagate with very different velocities, from one edge to the other, constitutes a clear signature of spin-charge separation, i.e.~a genuine peculiarity of 1D fermionic quantum systems.

Intuitively, the fact that the instantaneous edge-to-edge signal propagation occurs in the spin correlation function $S_j(t)$ is due to the edge modes being constituted of fermions with antiparallel spins for repulsive interactions ($U\!>\!0$). Due to the particle-hole symmetry inherent to the SSH model, a similar behavior can be observed for charge excitations when considering the case of attractive interactions ($U\!<\!0$):~indeed, in that case, a quasi-instantaneous edge-to-edge response is observed in the $C_L(t)$ signal.  

It should be noted that the non-interacting case ($U\!=\!0$) also displays quasi-instantaneous edge-to-edge correlation signal, however, due to the 4-fold degeneracy of the edge manifold, such a behavior is equally found in both correlation functions, $C_j$ and $S_j$. In this sense, the spin-charge separation identified above cannot be observed in the non-interacting regime. 

It is also worth to underline that the edge-to-edge correlation spreading does not occur in the pathological case where $\delta J\!=\!-J$, which is due to the vanishing correlation length of the edge states preventing any fluctuations in correlations.   
\begin{figure}
\includegraphics[scale=0.24]{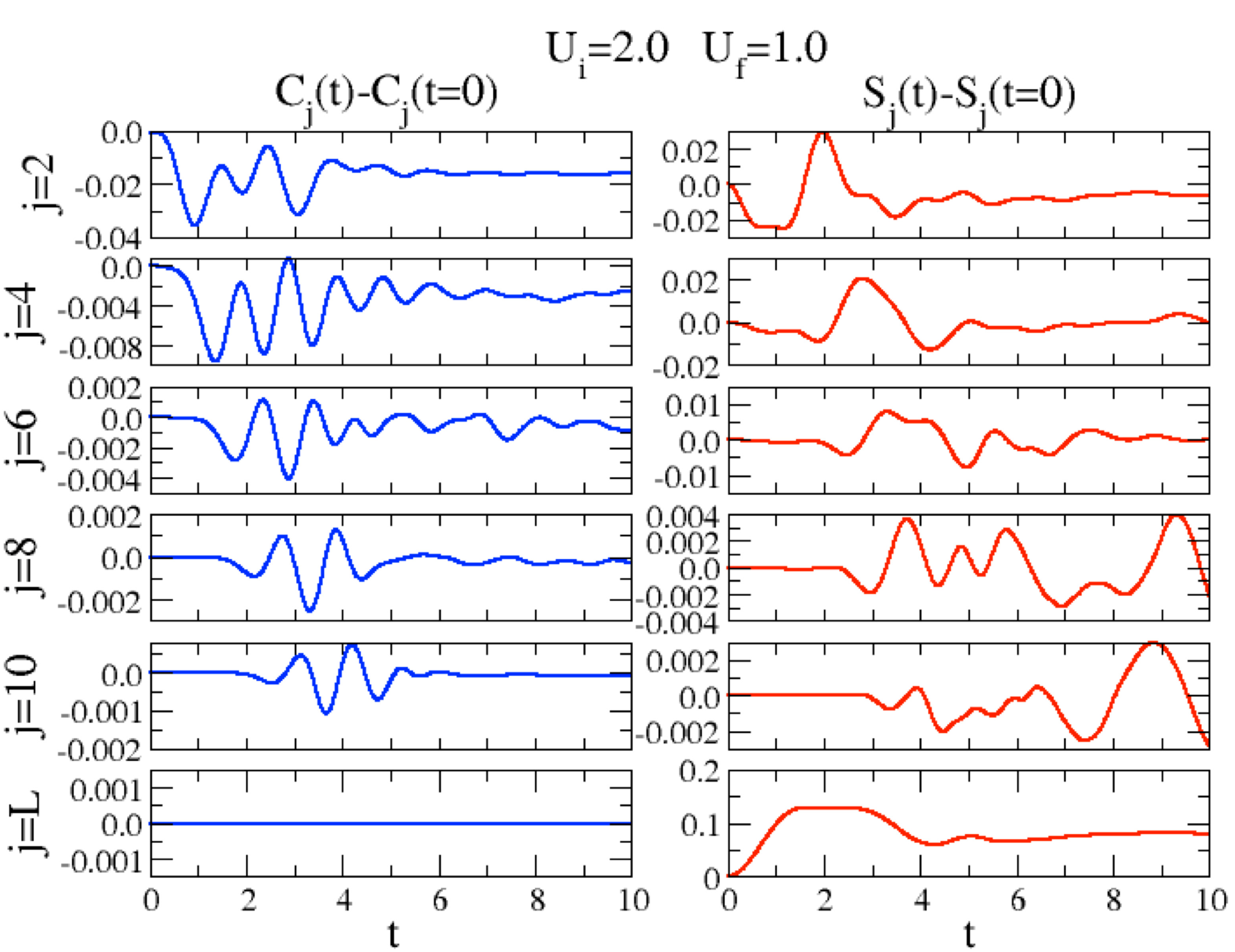}
\caption{(Color online) Spreading of the correlation $C_j(t)-C_j(t=0)$ and $S_j(t)-S_j(t=0)$ for $\delta J=-0.4$ and the interaction going suddenly from $U_i=2.0$ to $U_f=1.0$. All the results refer to a chain of length $L=40$ obtained by means of t-DMRG simulations keeping up to 512 DMRG states both for the static and for the dynamic and using time step $\delta t=0.01$}
\label{dynU}
\end{figure}

In order to verify that the quasi-instantaneous correlation signal is not an artifact attributed to the locality of the quench, we also analyze the correlation spreading that occurs upon subjecting the system to a global quench (i.e.~when the time-evolution is triggered by a sudden variation of a parameter defined in all lattice sites). One should note that this modification indeed generates a global deformation of the dispersion relation, whose shape is responsible for the strength of the propagation velocity~\cite{calabrese}. In the present SSH Hamiltonian, we propose to abruptly change the interaction strength from a certain initial value $U_i$ to a final value $U_f\!\ne\!U_i$; we note that a similar procedure can be obtained by varying $\delta J$. The result of this global quench is presented in Fig.~\ref{dynU}, which shows strikingly similar behavior as the one presented in Fig.~\ref{correDl} for the local quench:~a quasi-instantaneous edge-to-edge signal in $S_j$ only, while both correlations show light-cone-like correlation spreading in the bulk. We note that the edge-to-edge signal can be attributed to the fact that the interaction parameter $U$ affects the spatial localization of the edge states [Fig.~\ref{corre}(c)]:~a sudden variation in $U$ induces fluctuations at the edges, which due to the edge-to-edge entanglement, produces the quasi-instantaneous signal in $S_j$. This analysis implies that the non-trivial edge-to-edge signal is indeed an effect solely induced by the long-ranged correlations, i.e.~the entanglement, existing between the two (spatially-separated) edge states.\\ 

\paragraph{Discussion.}

This work studied the intriguing dynamical properties that emerge from the topological nature of the interacting fermionic Su-Schrieffer-Heeger model. We showed how a careful study of the correlation functions, which can be obtained through quasi-exact numerical methods, can reveal the nature of both the edge modes and that of the bulk dimerization, which characterize the interacting SSH model. Interestingly, we also revealed~\cite{SM} that the latter model features an exotic type of Peierls dimerization, where the spin gap dominates over the charge gap; this is in contrast with the usual dimerization associated with the extended Hubbard model~\cite{nakamura}. 

Importantly, we illustrated how the existence of long-distance entanglement, which stems from the topological nature of the system, could lead to strong consequences in the spreading of correlations upon a quench, including the possibility of observing an instantaneous edge-to-edge correlation signal. In particular, this suggests that the latter could be used as an experimental probe for entangled topological edge modes in cold-atom setups.

Furthermore, the spin-charge separation associated with edge-to-edge signals suggests that experimental realizations of the SSH~\cite{atala,meier} could offer a natural platform to study this genuine 1D effect in the laboratory. In particular, we point out that all the ingredients needed to explore these results experimentally are currently available in ultracold-atom setups; this includes methods to engineer the SSH model using optical superlattices~\cite{atala,meier}, the possibility of tuning inter-particle interactions in mixtures of ultracold fermions~\cite{jordens,hart,schreiber,cocchi,greif1}, as well as methods to probe both spin~\cite{greif,parson} and charge~\cite{endres} correlation functions.

We point out that while a box-like trapping potential~\cite{mukherjee} would facilitate the detection, properties associated with edge modes and light-cone propagation have been shown to be stable in the presence of the more standard harmonic confinement \cite{mazza,cheneau}. 

Finally, we anticipate that similar results could be obtained or generalized in other types of interacting topological systems with entangled edge modes, which opens intriguing perspective and motivates the search for novel realistic models with similar topological features.


\begin{acknowledgments}

{\it Acknowledgments:} Discussions with W. De Roeck, M. Di Liberto, E. Ercolessi, T. Giamarchi, G.I. Japaridze, C. V. Kraus, A. Montorsi, H. Pichler and F. Verstraete are acknowledged. 
L. B. and N. G. acknowledge ERC Starting Grant TopoCold for financial support. L. S. acknowledges support by the SFB 1227 "DQ-mat" of the German Research Foundation (DFG).  
\end{acknowledgments}

\section{Supplemental Material for:``\textit{Quenched dynamics and spin-charge separation in an interacting topological lattice}"}

\maketitle

\paragraph{Topological order in the presence of interaction.}
As shown in the main text, the presence of entangled degenerate edge modes is captured by the behavior of the correlation functions for the charge and spin degrees of freedom. In order to enforce our results, we hereby give further evidence for the existence of topological properties in the interacting SSH model, by analyzing other observables. For instance, a relevant signature associated with the existence of edge modes is provided by the even degeneracy in the lowest eigenvalues of the reduced density matrix~\cite{pollmann,turner,fidkowski,deng,kraus1}, usually called \emph{entanglement spectrum} (ES). In particular let us define:
\begin{equation}
\rho_A=\sum_{N,n}\lambda^N_n\rho^N_n , 
\end{equation}
where $A$ denotes a system bipartition and where $\rho^N_n$ describes a pure state of $N$ particles with the corresponding eigenvalues $\lambda^N_n$ (i.e.~the ES).
\begin{figure}
\includegraphics[scale=0.24]{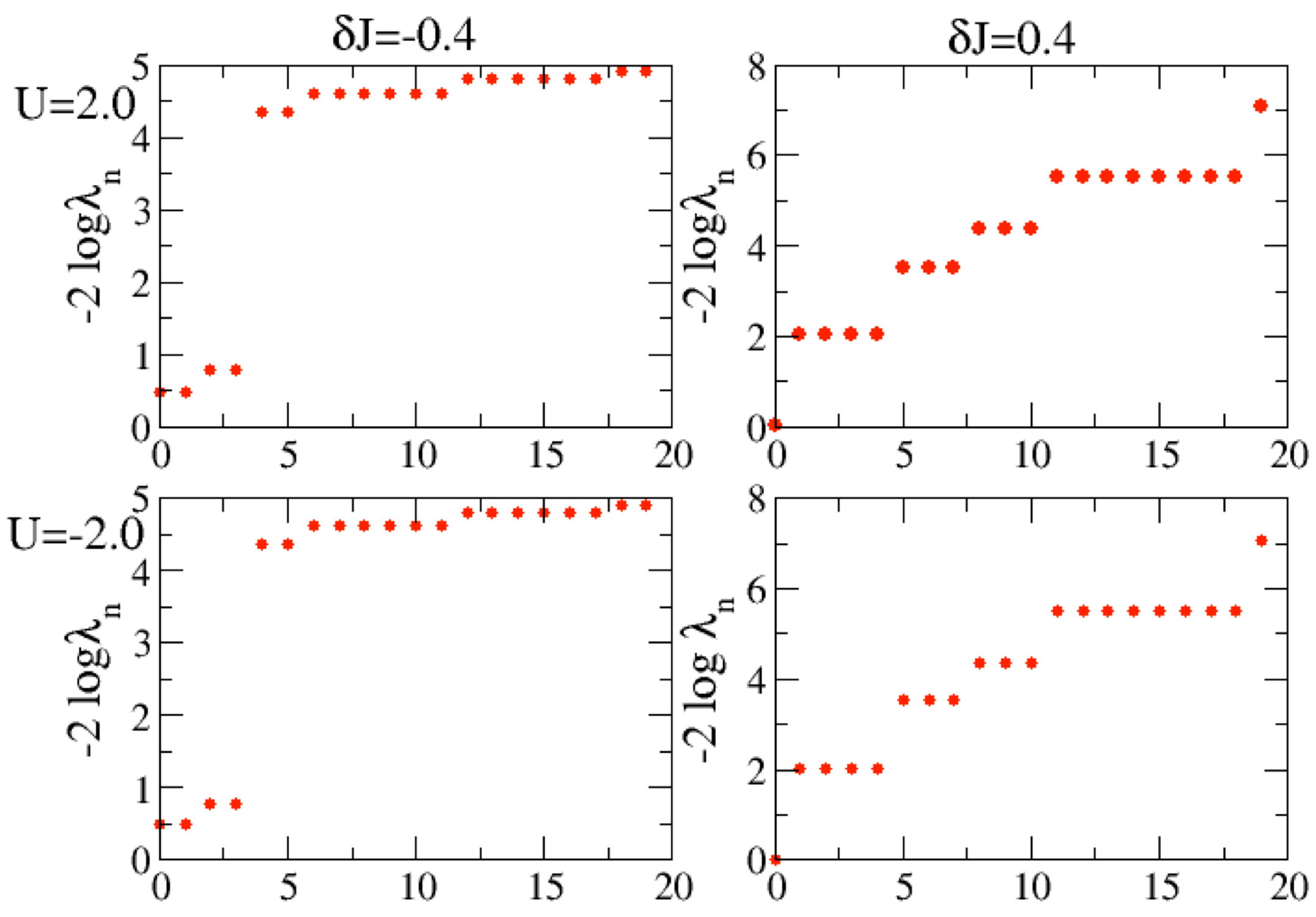}
\caption{(Color online) First 20 values of $\lambda_n^N$ in ascending order for different values of $U$ and $\delta J$. All the results refer to a chain of length $L=64$ obtained by means of DMRG simulations keeping up to 512 DMRG states and performing 5 finite size sweeps }
\label{es}
\end{figure}
By means of DMRG calculations, we confirm that the even degeneracy of the ES is present for $\delta J\!<\!0$, whereas it disappears once $\delta J$ becomes positive; see Fig. \ref{es}. We point out that the value of the degeneracy depends on the choice of boundary conditions (here we considered open boundary conditions with hard walls), however, this value should remain even within the non-trivial topological regime~\cite{ye}. Furthermore, the ES also signals certain symmetries of the Hamiltonian. Indeed, as clearly visible in Fig. \ref{es}, the ES has exactly the same values under the transformation $U\leftrightarrow-U$. This latter property reflects the particle-hole symmetry inherent to the interacting SSH fermionic model; we note that this symmetry is  absent in its bosonic counterpart.

In order to check the robustness of zero energy edge modes in the presence of finite interaction, an additional signature is given by the following quantity:
\begin{equation}
\Delta n_i=n_i^{(N+1)}-n_i^{(N)} ,
\label{dn}
\end{equation}
where $n_i^{(N)}$ is the occupation number at lattice site $i$ in a system with $N$ particles; the observable $\Delta n_i$ describes how an added particle distributes itself in the system under scrutiny.
\begin{figure}
\includegraphics[scale=0.24]{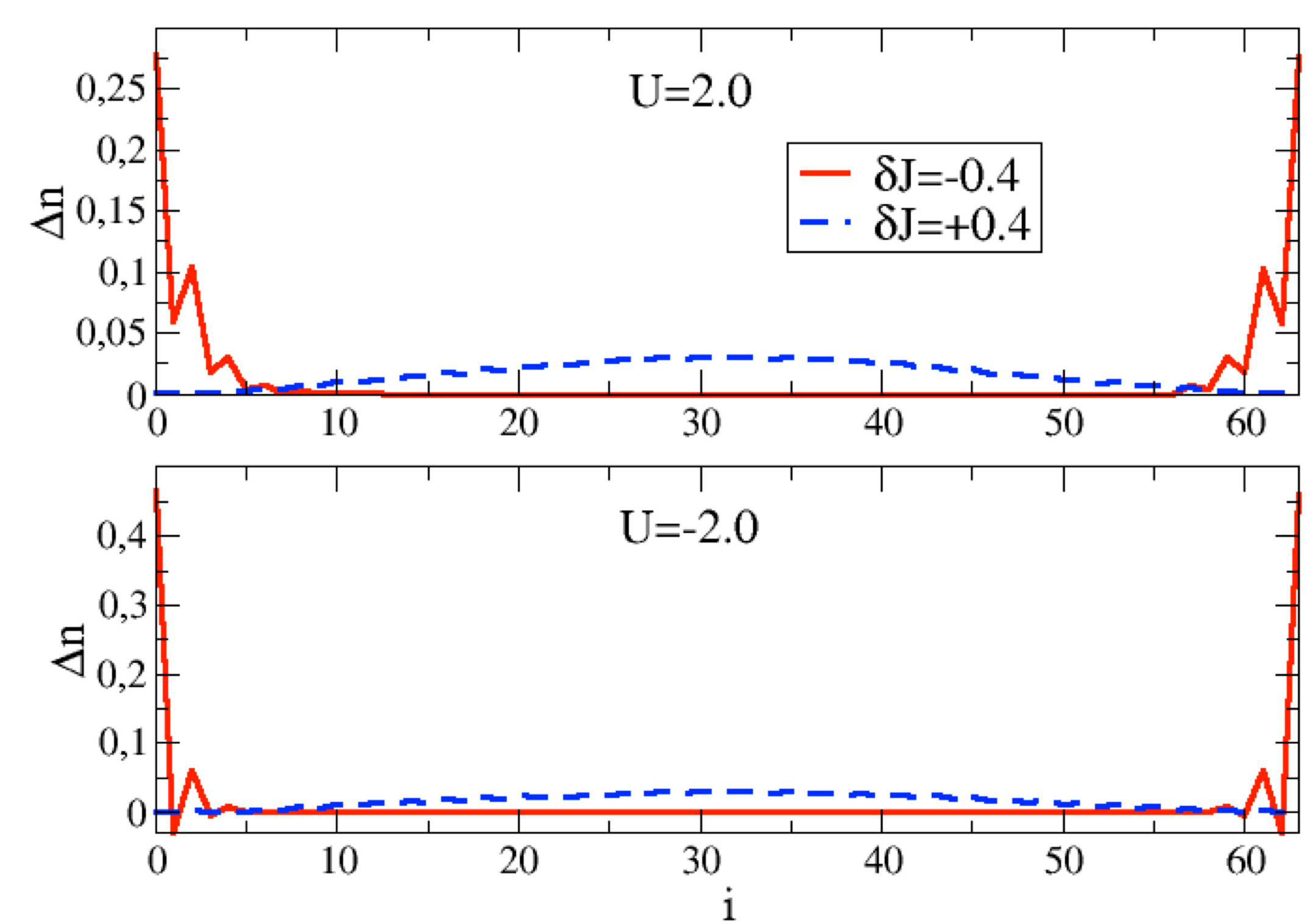}
\caption{(Color online) \textit{upper panel)}: $\Delta n$ for two different values of $\delta t$ and $U=2.0$ 
\textit{lower panel)}: $\Delta n$ for two different values of $\delta J$ and $U=-2.0$. All the result refer to a chain of length $L=64$ obtained by means of DMRG simulations keeping up to 512 DMRG states and performing 5 finite size sweeps }
\label{dn}
\end{figure}
As clearly visible in Fig.~\ref{dn}, our DMRG simulations shows that the extra particle always accumulates in the system edges, whenever $\delta J<0$; in particular, this is observed for both repulsive and attractive interactions. This last result can be attributed to the fact that degenerate zero energy edge modes are present in a topological phase, and thus, that the extra particle is free to occupy the more external sites, with a zero energy cost. On the other hand, a positive $\delta J$ forces the extra particle to distribute itself in the center of the bulk of the lattice; here the extra fermion always has to pay a finite energy, which is minimized at the center of system where the kinetic energy can be maximized. 

It is also interesting to notice that the behavior of both observables [Eqs.~\eqref{es} and \eqref{dn}] remains unchanged in the non-interacting limit $U\!=\!0$. Indeed, as we show in the main text, the effect of the interaction consists in reducing the ground-state degeneracy (from 4 to 2), which is then reflected in a peculiar spin and charge separation captured by the correlation functions analyzed in the main text.
\paragraph{Peierles dimerization.}
Peierles dimerization (PD), also called bond-ordering wave, is a quantum regime where particles form bonds in alternating sites, hence reflecting a spontaneous reflection-symmetry breaking. In the SSH model, this phase takes place for any value of $\delta J$:~however, while dimerization appears in all lattice sites for $\delta J\!>\!0$, this effect only occurs in the bulk for $\delta J\!<\!0$. In the context of 1D fermionic systems, bosonization predicts the presence of PD in the extended Hubbard model~\cite{nakamura}. It has also been shown~\cite{barbiero} that the order parameters for such a phase are the parity operators defined in the charge and spin sectors,
\begin{equation}
O_p^C(j)=e^{i \pi\sum_{l=0}^jS_l^C},
\label{pc}
\end{equation} 
\begin{equation}
O_p^S(j)=e^{i \pi\sum_{l=0}^jS_l^S},
\label{ps}
\end{equation}
\begin{figure}
\includegraphics[scale=0.20]{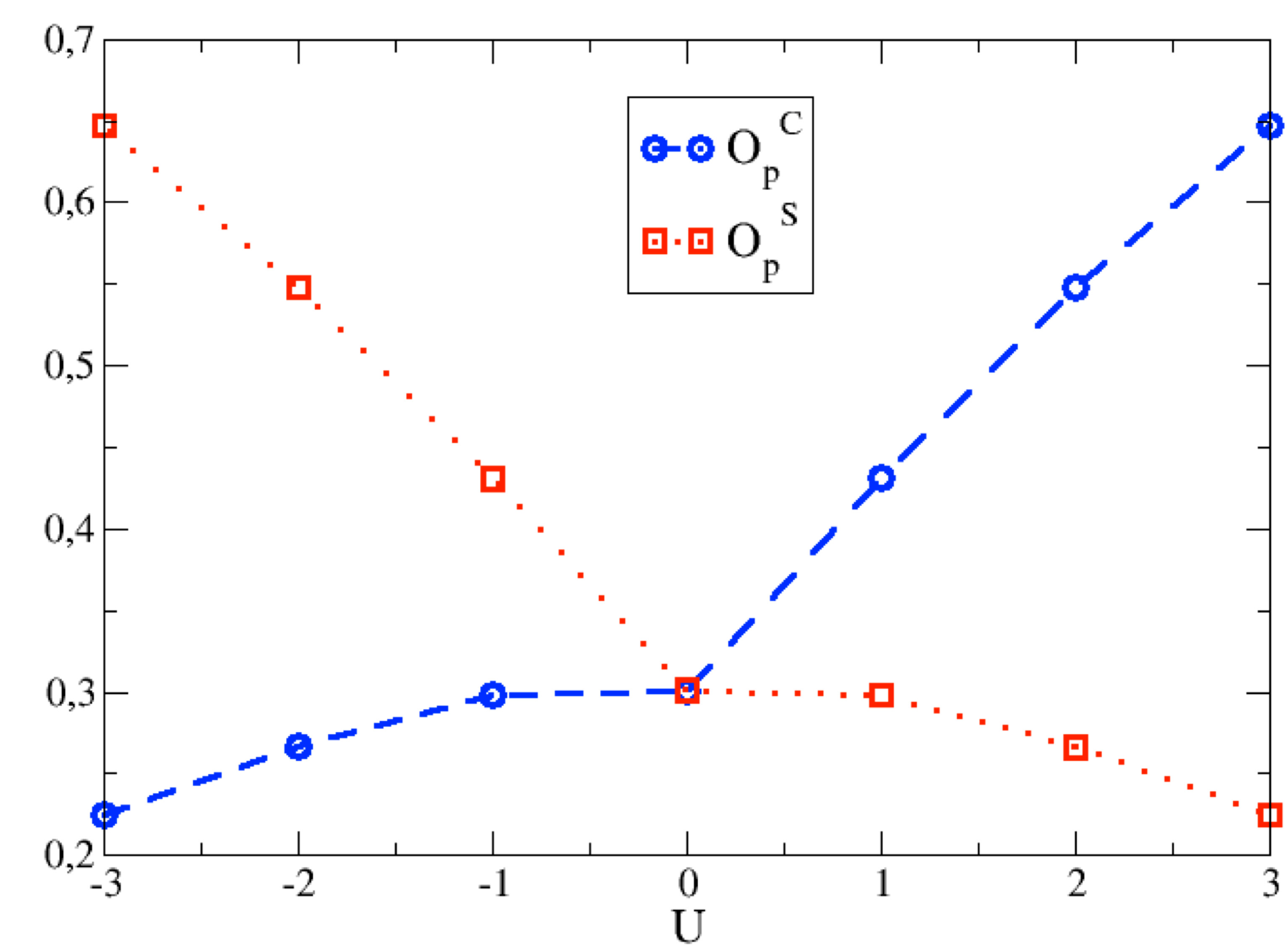}
\caption{(Color online) $O_p^C$ and $O_p^S$ as a function of $U$ for a system of length $L=64$. All the results are obtained by means of DMRG simulations keeping up to 512 DMRG states and performing 5 finite size sweeps }
\label{parity}
\end{figure}
where $S_l^C\!=\!(n_l-1)$ and $S_l^S\!=\!(n_{\uparrow,l}-n_{\downarrow,l})$, as introduced in the main text. A finite value of both observables [Eqs.~\eqref{pc}) and \eqref{ps}] signals a regime where both charge and spin gaps are present, which is a peculiar feature of PD in fermionic systems. As in the case of the extended Hubbard model (EHM), $O_p^C\!>\!O_p^S$ indicates that the charge gap dominates over the spin gap; this implies that the bonds in the system are formed by two fermions with opposite spins. Due to the absence of particle-hole symmetry, the EHM does not support the opposite scenario, namely, $O_p^C\!<\!O_p^S$, where dimerization would occur between pairs and holons.
As visible in Fig. \ref{parity}, the latter scenario is however possible in the attractive regime of the SSH model (for $U\!<\!0$):~since $O_p^C\!<\!O_p^S$, this model does exhibit a pair-holon dimerization. On the other hand, for $U\!>\!0$, a PD similar to the one occurring in the EHM takes place. It is also relevant to notice that the 4-fold degeneracy associated with the $U\!=\!0$ case allows for both types of PD, as sucggested by the equality $O_p^C\!=\!O_p^S$ .
\paragraph{Conclusions.}
In this supplemental material we enforced our results regarding the persistence of a topological order in the Su-Schrieffer-Heeger model for finite interaction. Moreover we precisely characterized the Peierles dimerization occurring in the aforementioned model in different interacting regimes. This last point has allowed us to reveal two non-trivial types of dimerization (for attractive and vanishing interactions) which are not present in the extended Hubbard model.


\begin{thebibliography}{99}

\bibitem{hasan} M. Z. Hasan and C. L. Kane, Rev. Mod. Phys. \textbf{82}, 3045
(2010).

\bibitem{qi} Xiao-Liang Qi, Shou-Cheng Zhang, Rev. Mod. Phys. \textbf{83}, 1057-
1110 (2011).

\bibitem{Asorey} M. Asorey, Nat. Phys. {\bf 12}, 616 (2016).

\bibitem{Goldman_topology} N. Goldman, J. C. Budich and P. Zoller, Nature Physics {\bf 12}, 639 (2016).

\bibitem{Ozawa_review} T. Ozawa \emph{et al.}, 	arXiv:1802.04173.

\bibitem{kane} C.L. Kane and E.J. Mele, Phys. Rev. Lett. \textbf{95}, 146802 (2005).

\bibitem{bernevig} B.A. Bernevig, and S.C. Zhang, Phys. Rev. Lett. \textbf{96}, 106802
(2006); B.A. Bernevig, T.L. Hughes, S.-C. Zhang, Science \textbf{314},
1757-1761 (2006).

\bibitem{klitzking} K. Von Klitzing, G. Dorda, M. Pepper, Phys. Rev. Lett. \textbf{45}, 494-497 (1980).

\bibitem{laughlin} R. B. Laughlin, Phys. Rev. B \textbf{23}, 5632-5633 (1981).

\bibitem{thouless} D.J. Thouless, M. Kohmoto, M.P. Nightingale, and M. den Nijs,
Phys. Rev. Lett. \textbf{49}, 405 (1982).

\bibitem{moore} G. Moore, and N. Read, Nuclear Physics B \textbf{360}, 362-396
(1991).

\bibitem{kitaev} Kitaev, Physics-Uspekhi 44, 131 (2001).

\bibitem{ryu} Y. Ryu and S. Hatsugai Phys. Rev. Lett. \textbf{89}, 077002 (2002).

\bibitem{Jiang_2011} L. Jiang \emph{et al.}, Phys. Rev. Lett. {\bf 106}, 220402 (2011).

\bibitem{Alicea_2011} J. Alicea, Y. Oreg, G. Refael, F. von Oppen and M. P. A. Fisher, Nature Physics {\bf 7}, 412 (2011).

\bibitem{deplace} P. Delplace, D. Ullmo, G. Montambaux, Phys. Rev. B \textbf{84},
195452 (2011).

\bibitem{kraus} Y. E. Kraus, Y. Lahini, Z. Ringel, M. Verbin and O. Zilberberg,
Phys. Rev. Lett. \textbf{109}, 106402 (2012).

\bibitem{verbin} M. Verbin, O. Zilberberg, Y. E. Kraus, Y. Lahini and Y. Silberberg,
Phys. Rev. Lett. \textbf{110}, 076403 (2013).

\bibitem{Christina_Marcello} C. V. Kraus, M. Dalmonte, M. A. Baranov, A. M. L\"auchli, and P. Zoller, Phys. Rev. Lett. {\bf 111}, 173004 (2013).

\bibitem{diliberto} M. Di Liberto, A. Recati, I. Carusotto, and C. Menotti, Phys. Rev. A {\bf 94}, 062704 (2016).

\bibitem{gorlach} M. A. Gorlach, A. N. Poddubny Phys. Rev. A {\bf 95}, 053866 (2017).

\bibitem{Alejandro} J. J\"unemann, A. Piga, S.-J. Ran, M. Lewenstein, M. Rizzi, and A. Bermudez, Phys. Rev. X {\bf 7}, 031057  (2017).

\bibitem{icfo_dynamical} D. Gonzalez-Cuadra, P. R. Grzybowski, A. Dauphin and M. Lewenstein, arXiv:1802.05689.

\bibitem{sbierski} B. Sbierski, and C. Karrasch, arXiv:1805.00839.


%
%
%
%

\bibitem{su} W. P. Su, J. R. Schrieffer, and A. J. Heeger, Phys. Rev.
Lett. \textbf{42}, 1698 (1979).

\bibitem{heeger} A. J. Heeger, S. Kivelson, J. R. Schrieffer, and W. P. Su, Rev. Mod. Phys. \textbf{60}, 781 (1988).

\bibitem{jackiw} R. Jackiw and C. Rebbi, Phys. Rev. D \textbf{13}, 3398 (1976).

\bibitem{atala} M. Atala, M. Aidelsburger, J. T. Barreiro, D. Abanin, T. Kitagawa, E. Demler, I. Bloch Nature Physics \textbf{9}, 795-800 (2013).

\bibitem{meier} E. J. Meier, F. A. An, B. Gadway, Nature Communications textbf{7}, 13986 (2016).

\bibitem{Malkova2009} N. Malkova, I. Hromada, X. Wang, G. Bryant, and Z. Chen, Optics letters {\bf 34} , 1633 (2009).

\bibitem{stjean} P. St-Jean, V. Goblot, E. Galopin, A. Lema"tre, T. Ozawa, L. Le Gratiet, I.Sagnes, J. Bloch, A. Amo, Nature Photonics {\bf 11}, 651 (2017).

\bibitem{giamarchi} T. Giamarchi, Quantum Physics in One Dimension (Oxford Univ. Press, Oxford, 2004).

\bibitem{tomonaga} S. Tomonaga, Progress in Theoretical Physics, \textbf{5}, 544 (1950).

\bibitem{luttinger} J. M. Luttinger, Journal of Mathematical Physics, \textbf{4}, 1154 (1963).

\bibitem{kim} B. J. Kim, H. Koh, E. Rotenberg, S.-J. Oh, H. Eisaki, N. Motoyama, S. Uchida, T. Tohyama, S. Maekawa, Z.-X. Shen, and C. Kim, Nat. Phys. \textbf{2} 397Ð401 (2006).

\bibitem{jompol} Y. Jompol, C. J. B. Ford, J. P. Griffiths, I. Farrer, G. A. C. Jones, D. Anderson, D. A. Ritchie, T. W. Silk, A. J. Schofield, Science \textbf{325} (2009) 597-601. 

\bibitem{lieb} E. H. Lieb and D. W. Robinson, Comm. Math. Phys. \textbf{28}, 251 (1972).

\bibitem{nachtergaele} B. Nachtergaele and R. Sims, Commun. Math. Phys. \textbf{265}, 119 (2006).

\bibitem{hastings1} M. B. Hastings, Phys. Rev. B, \textbf{69}, 104431 (2004).

\bibitem{hastings2} M. B. Hastings and T. Koma, Commun. Math. Phys., \textbf{265}, 781 (2006).

\bibitem{eisert} J. Eisert, M. Cramer, and M. B. Plenio, Rev. Mod. Phys., \textbf{82}, 277 (2010).

\bibitem{rigol} M. Rigol, V. Dunjko, V. Yurovsky, and M. Olshanii, Phys. Rev. Lett. \textbf{98}, 050405 (2007). 
 
\bibitem{gong} X.-Z. Gong, and L. M. Duan,  New J. Phys. 15, 113051 (2013).

\bibitem{bose} S. Bose, Contemp. Phys. \textbf{48}, 13Ð30 (2007).

\bibitem{hastings3} M. Hastings, J. Stat. Mech. \textbf{2007}, P08024 (2007).

\bibitem{bravyi} S. Bravyi, M. B. Hastings, and F. Verstraete, Phys. Rev. Lett. 97, 050401 (2006).

\bibitem{cheneau} M. Cheneau, P. Barmettler, D. Poletti, M. Endres, P. Schauss, T. Fukuhara, C. Gross, I. Bloch, C. Kollath, and S. Kuhr, Nature \textbf{481}, 484 (2012).

\bibitem{langen} T. Langen, R. Geiger, M. Kuhnert, B. Rauer, and J. Schmiedmayer, Nat. Phys. \textbf{9}, 640 (2013).






\bibitem{richerme} P. Richerme, Z.-X. Gong, A. Lee, C. Senko, J. Smith, M. Foss-Feig, S. Michalakis, A. V. Gorshkov, C. Monroe, Nature 511, 198 (2014)

\bibitem{jurcevic} P. Jurcevic, B. P. Lanyon, P. Hauke, C. Hempel, P. Zoller, R. Blatt, C. F. Roos, Nature 511, 202 (2014)


\bibitem{Caio2015} M. D. Caio, N. R. Cooper and M. J. Bhaseen, Phys. Rev. Lett.
{\bf 115} 236403 (2015).

\bibitem{Alessio2015} L. DÕAlessio and M. Rigol, Nat. Commun. {\bf 6} 8336 (2015).

\bibitem{strinati} M. Calvanese Strinati, L. Mazza, M. Endres, D. Rossini, R. Fazio, Phys. Rev. B \textbf{94}, 024302 (2016).

\bibitem{sacramento} P. D. Sacramento, Phys. Rev. E \textbf{93}, 062117 (2016).


\bibitem{Caio2016} M. D. Caio, N. R. Cooper and M. J. Bhaseen, Phys. Rev. B
{\bf 94} 155104 (2016).

\bibitem{Hu2016} Y. Hu, P. Zoller and J. C. Budich, Phys. Rev. Lett. {\bf 117} 126803 (2016).

\bibitem{Wilson2016} J. H. Wilson, J. C. W. Song and G. Refael, Phys. Rev. Lett.
{\bf 117} 235302 (2016).

\bibitem{Dehghani2016} H. Dehghani and A. Mitra, Phys. Rev. B {\bf 93} 205437 (2016).

\bibitem{Unal2016} F. Nur \"Unal, E. J. Mueller and M. \"O. Oktel, Phys. Rev. A
{\bf 94} 053604 (2016).

\bibitem{Wang2017} C. Wang, P. Zhang, X. Chen, J. Yu and H. Zhai, Phys. Rev. Lett. {\bf 118}, 185701 (2017).

\bibitem{Flaschner2017} N. Fl\"aschner, D. Vogel, M. Tarnowski, B. S. Rem, D.-S. L\"uhmann, M. Heyl, J. C. Budich, L. Mathey, K. Sengstock, C. Weitenberg, Nature Physics (2017).

\bibitem{Tarnowski2017} M. Tarnowski, F. N. \"Unal, N. Fl\"aschner, B. S. Rem, A. Eckardt, K. Sengstock, C. Weitenberg, arXiv:1709.01046.

\bibitem{grusdt} F. Grusdt, M. Hšning, M. Fleischhauer, Phys. Rev. Lett. \textbf{110}, 260405 (2013).

\bibitem{wang} D. Wang, S. Xu, Y. Wang, C. Wu, Phys. Rev B \textbf{91}, 115118 (2015).

\bibitem{ye} B. Ye, L. Mu,  H. Fan, Phys. Rev. B \textbf{94}, 165167 (2016).

\bibitem{peierls} R. Peierls, \textit{Surprises in Theoretical Physics (Princeton University Press}, Princeton, 1979).

\bibitem{yu} W. C. Yu, Y. C. Li, P. D. Sacramento, H.-Q. Lin, Phys. Rev. B \textbf{94}, 245123 (2016).

\bibitem{barbiero} L. Barbiero, A. Montorsi, and M. Roncaglia, Phys. Rev. B \textbf{88}, 035109 (2013).

\bibitem{SM} see Supplementary Material

\bibitem{pollmann} F. Pollmann, E. Berg, A. M. Turner, M. Oshikawa, Phys. Rev. B \textbf{81}, 064439 (2010); A. M. Turner, F. Pollmann, E. Berg, Phys. Rev. B \textbf{83}, 075102 (2011); L. Fidkowski and A. Kitaev, Phys. Rev. B \textbf{83}, 075103 (2011); X. Deng and, L. Santos, Phys. Rev. B \textbf{84}, 085138 (2011).

\bibitem{white} S. R. White, Phys. Rev. Lett. {\bf 69}, 2863 (1992).

\bibitem{nakamura} M. Nakamura, Phys. Rev. B \textbf{61}, 16377 (2000).

\bibitem{campos} L. Campos Venuti, C. Degli Esposti Boschi, M. Roncaglia, Phys. Rev. Lett. \textbf{96}, 247206 (2006).

\bibitem{white1} S. R. White and A. E. Feiguin, Phys. Rev. Lett. \textbf{93}, 076401 (2004);
A. E. Feiguin and S. R. White, Phys. Rev. B \textbf{72}, 020404(R) (2005).

\bibitem{note} We checked that a different choice of the local perturbation does not affect the behavior of the correlation spreading.

\bibitem{calabrese} P. Calabrese and J. Cardy, Phys. Rev. Lett. \textbf{96}, 136801 (2006).

\bibitem{jordens} R. J\"ordens, N. Strohmaier, K. G\"unter, H. Moritz, T. Esslinger, Nature (London) \textbf{455}, 204-207 (2008).

\bibitem{hart} R. A. Hart, P. M. Duarte, T.-L. Yang, X. Liu, T. Paiva, E. Khatami, R. T. Scalettar, N. Trivedi, D. A. Huse, R. G. Hulet, Nature \textbf{519}, 211-214 (2015).

\bibitem{schreiber} M. Schreiber, S. S. Hodgman, P. Bordia, H. P. L\"uschen, M. H. Fischer, R. Vosk, E. Altman, U. Schneider, I. Bloch, Science \textbf{349}, 842 (2015).

\bibitem{cocchi} E. Cocchi, L. A. Miller, J. H. Drewes, M. Koschorreck, D. Pertot, F. Brennecke, M. K\"ohl, Phys. Rev. Lett. \textbf{116}, 175301 (2016).

\bibitem{greif1} D. Greif, M. F. Parsons, A. Mazurenko, C. S. Chiu, S. Blatt, F. Huber, G. Ji, M. Greiner, Science \textbf{351}, 953-957 (2016).

\bibitem{greif} D. Greif, T. Uehlinger, G. Jotzu, L. Tarruell, T. Esslinger, Science \textbf{340}, 1307-1310 (2013).

\bibitem{parson} M. F. Parsons, A. Mazurenko, C. S. Chiu, G. Ji, D. Greif, M. Greiner, Science \textbf{353}, 1253-1256 (2016).

\bibitem{endres} M. Endres, M. Cheneau, T. Fukuhara, C. Weitenberg, P. Schauss, C. Gross, L. Mazza, M.C. Banuls, L. Pollet, I. Bloch, S. Kuhr, Science \textbf{334}, 200 (2011).

\bibitem{mukherjee} B. Mukherjee, Z. Yan, P. B. Patel, Z. Hadzibabic, T. Yefsah, J. Struck, and M. W. Zwierlein, Phys. Rev. Lett. \textbf{118}, 123401 (2017). 

\bibitem{mazza} L. Mazza, M. Aidelsburger, H.-H. Tu, N. Goldman, M. Burrello, New J. Phys. 17 (2015) 105001.

\end{thebibliography}

\begin{thebibliography}{99}

\bibitem{pollmann} F. Pollmann, E. Berg, A. M. Turner, M. Oshikawa, Phys. Rev. B \textbf{81}, 064439 (2010).

\bibitem{turner} A. M. Turner, F. Pollmann, E. Berg, Phys. Rev. B \textbf{83}, 075102 (2011).

\bibitem{fidkowski} L. Fidkowski and A. Kitaev, Phys. Rev. B \textbf{83}, 075103 (2011).

\bibitem{deng} X. Deng and, L. Santos, Phys. Rev. B \textbf{84}, 085138 (2011).

\bibitem{kraus1} C. V. Kraus, M. Dalmonte, M. A. Baranov, A. M. Laeuchli, P. Zoller, Phys. Rev. Lett. \textbf{111}, 173004 (2013).

\bibitem{wang} D. Wang, S. Xu, Y. Wang, C. Wu, Phys. Rev B \textbf{91}, 115118 (2015).

\bibitem{ye} B. Ye, L. Mu,  H. Fan, Phys. Rev. B \textbf{94}, 165167 (2016).

\bibitem{nakamura} M. Nakamura, Phys. Rev. B \textbf{61}, 16377 (2000).

\bibitem{barbiero} L. Barbiero, A. Montorsi, and M. Roncaglia, Phys. Rev. B \textbf{88}, 035109 (2013).


\end{thebibliography}
\end{document}